\documentclass[modern]{aastex61}

\usepackage{graphicx}
\usepackage{amsmath}
\usepackage{amssymb}
\usepackage{enumitem}

\newcommand\mybar{\kern1pt\rule[-\dp\strutbox]{.8pt}{\baselineskip}\kern1pt}

\setlist[itemize]{noitemsep, topsep=0pt, leftmargin=*}

\shorttitle{PTA Noise from Sun's Brownian Motion}
\shortauthors{Loeb}

\begin{document}

\title{Pulsar Timing Noise from Brownian Motion of the Sun}

\author{Abraham Loeb}
\affiliation{Astronomy Department, Harvard University, 60 Garden
  St., Cambridge, MA 02138, USA}

\begin{abstract}
Recently, Pulsar Timing Arrays (PTAs) reported a signal at nanohertz
frequencies consistent with a stochastic gravitational wave
background. Here, I show that the Brownian motion of the Sun as a
result of its random gravitational interactions with the cluster of
thousands of unmodeled Main-belt asteroids of diameters $\lesssim
80$~km, combined with small inaccuracies in the estimated masses of
larger asteroids, introduces timing noise for pulsars which is up to
$20\%$ of the reported signal at frequencies of $\sim 1/({\rm
  few~years})$.  The asteroid contribution needs to be modeled better
in order to obtain accurate inferences from the PTA signal.
\end{abstract}

\section{Introduction}

Pulsar Timing Arrays (PTAs) monitor a population of millisecond radio
pulsars~\citep{Foster1990} in an attempt to detect correlated shifts
in the arrival times of their radio
pulses~\citep{Hellings1983,Finn2009}. Several PTAs reported recently
the discovery of a correclated signal at frequencies of 1-10
nanohertz~\citep{Ant2022,Agazie2023,Reardon2023}. The signal was
interpreted as a stochastic gravitational wave background, potentially
from mergers of supermassive black hole binaries throughout cosmic
history~\citep{Agaz23,PadL}.

Here, I show that the detected stochastic signal is contaminated by
the Brownian motion of the Sun as a result of its interactions with
unmodeled Main-belt asteroids of diameters $\lesssim 80$~km, as well
as mass inaccuracies for larger asteroids. 

The PTAs identification of a stochastic gravitational wave
background~\citep{Ant2022,Agazie2023,Reardon2023} used
the~\citet{Hellings1983} angular correlations among pulsars, based on
the quadrupolar nature of gravitational waves. The Doppler noise
considered here is expected to produce dipolar angular correlations
among pulsars. However, given the small number of independent
correlation times ($\sim 5$ periods of $\sim 3$ years) available
during the 15 years of PTA observations, it is challenging to separate
the 3D random walk of a dipole (with a $\cos \theta$ angular
correlation of the pulsars) sourced by asteroids around the Sun from a
quadrupolar random walk sourced by a stochastic gravitational wave
background; the resultant high uncertainty is evident from the large
error bars in the Hellings-Down correlation data plotted in Fig. 1c
of~\citet{Agazie2023}.  Another distinguishing feature is that the
asteroid noise is constrained to the orbital period range of 3-10~yr,
whereas the gravitational wave background has a wider range. Finally,
because the Main belt asteroids lie within 30 degrees of the ecliptic
plane, which is inclined relative to the Galactic plane, the Brownian
motion that they induce could be separated from a stochastic
gravitational wave background.

In \S~2, I derive the amplitude and frequencies of the resulting
Doppler noise in pulsar timing, and in \S~3,~I discuss future
prospects for characterizing this unmodeled noise.

\section{Brownian Motion of the Sun}

The Brownian motion of a massive object embedded in a gravitationally
bound cluster of low-mass perturbers was studied both analytically and
using N-body simulations in the context of massive black holes in star
clusters~\citep{Chatterjee1,Chatterjee2,Merritt}. Here, I apply these
results to the Solar system, where the Sun is embedded in a cluster of
unmodeled Main-belt asteroids~\citep{Nes1,Nes2}. The many-body system
of asteroids with diameters $\lesssim 80$km represent objects that are
not included in the Solar system ephemeris~\citep{Eph} and is known to
have a substantial spread in inclinations and
eccentricities~\citep{Davis,Maeda}.

Previous analysis of PTA data was based on the Solar system ephemeris
calculated from well-studied objects in the Solar
system~(see~\citet{Agnoise}~and~\citet{Noise2}, and references
therein). The Solar system ephemeris model includes 343
asteroids~\citep{Eph}, but according to the cumulative asteroid number
versus diameter in \citet{Bottke}, there are many more than 343 Main
belt asteroids below a mass scale of $\sim 10^{21}~{\rm g}$.

\citet{Caballero2018} derived limits on individual unmodeled objects
with masses as small as $\sim 10^{-11}M_\odot=2\times 10^{22}~{\rm g}$
in Keplerian orbits around the Sun. However, a cluster of unmodeled
perturbers which are not included in the Solar system
ephemeris~\citep{Eph}, is known to exist in the Main asteroid
belt~\citep{Davis,Maeda,Nes1,Nes2}. These unmodeled objects, which are
not part of the Solar system ephemeris~\citep{Eph}, result in a
Brownian motion of the Sun and introduce a stochastic noise to the PTA
signal that cannot be picked up through a fit to a Keplerian orbit of
a single unmodeled object, as done
by~\citet{Caballero2018}~or~\citet{Guo1,Guo2}. Below I derive the
characteristic amplitude and frequencies of this unmodeled noise based
on known properties of the Main asteroid
belt~\citep{Davis,Bottke,Maeda,Nes1,Nes2}.

The motion of the Sun relative to the Solar system barycenter
satisfies momentum conservation with all objects of masses $m_i$ and
velocity ${\bf v}_i$ in the Solar system,
\begin{equation}
M_\odot {\bf v}_\odot = \Sigma_{i} m_i {\bf v}_i~.
\label{mom}
\end{equation}
where ${\bf v}_\odot$ is the Sun's velocity. The effects of the
planets, minor planets, moons and massive asteroids is already
included in the ephemeris model of the Solar system used by
PTAs~\citep{Eph,Agnoise,Noise2}.

Here, I consider the large population of unmodeled asteroids
($\lesssim 80$km in diameter) as a source of residual noise in
correlated pulsar timing (see also~\citet{Fedd}). Squaring both sides
of equation~(\ref{mom}), taking a statistical average (denoted
hereafter by angular brackets, following the ergodic theorem) and
ignoring velocity correlations among the many asteroids, I get the
dispersion in random velocity fluctuations of the Sun as a result of
its stochastic gravitational interactions with the cluster of
unmodeled asteroids,
\begin{equation}
\langle \vert v_\odot^2 \vert \rangle = {N_a m_a^2\over M_\odot^2}\langle
\vert v_a^2\vert\rangle ~,
\label{twomom}
\end{equation}
where $m_a$ and $v_a$ are the characteristic mass and velocity of the
unmodeled asteroids and $N_a\gg 1$ is their total number.

As expected, equation~(\ref{twomom}) yields kinetic-energy equipartition
between the central mass and its perturbers if the total mass of the
perturbers equals the central mass, $N_a m_a= M_\odot$. However, for a
smaller perturber population, as in the case of the Solar system,
Poisson ($\sim \sqrt{N_a}$) fluctuations yield a reduced velocity
dispersion for the central mass.

Main-belt asteroids with a typical mass $m\lesssim 10^{-12}M_\odot$ and a
characteristic speed $\sim 20~{\rm km~s^{-1}}$, yield a Brownian
motion of the Sun with a velocity dispersion of order,
\begin{equation}
\delta v \equiv \langle \vert v_\odot^2 \vert \rangle^{1/2} = 3.2\times
10^{-5}~{\rm cm~s^{-1}}\left({N_a\over 
  10^3}\right)^{1/2}\left({m_a\over 10^{21}~{\rm g}}\right)\left({\langle \vert v_a^2\vert \rangle^{1/2}\over 20~{\rm km~s^{-1}}}\right)~.
\label{nomom}
\end{equation}

Assuming a median mass density for asteroids of $\sim 3$~g~cm$^{-3}$,
the mass of a spherical asteroid with a diameter $D$ is,
$m_a(D)=0.8\times 10^{21}~{\rm g}(D/80~{\rm km})^3$.  Adopting the
differential number per diameter distribution of Main belt
asteroids~\citep{Bottke}, $dN_a/dD=10~{\rm km}^{-1}(D/80~{\rm
  km})^{-3.5}$ for $D\gtrsim 8$~km, 
\begin{equation}
(\delta v)^2 =\frac{\langle\vert v_a^2 \vert\rangle}{M_\odot^{2}}\int_{\rm 8~km}^{\rm 80~km} \frac{dN_a}{dD}m_{a}^2(D)dD
  =(1.2\times 10^{-5}~{\rm cm}~{\rm s}^{-1})^2\left({\langle \vert
    v_a^2\vert \rangle^{1/2}\over 20~{\rm km~s^{-1}}}\right)^2~.
  \label{newo}
\end{equation}

A fractional uncertainty $\epsilon$ in the masses of asteroids with
$D>80~{\rm km}$ up to a maximum diameter of $D_{\rm max}$, gives a
correction factor of $F=\{1+\epsilon^2[(D_{\rm max}/80~{\rm
    km})^{3.5}-1]\}^{1/2}$ in the value of $\delta v$ in
equation~(\ref{newo}).  Based on the mass uncertainties listed
in~\cite{Baer1,Baer2}, $F\approx 1.4$. However, there are additional
modeling errors as a result of spacecraft tracking
inaccuracies~\citep{Fienga,Kuchynka,Noise2} and chaotic behavior of
the Solar system~\citep{Laskar,Souami,Zeebe}. To effectively
incorporate these uncertainties, we adopt $\epsilon \sim 10$\% and
$D_{\rm max}\sim 500$~km, which yields $F\sim 2.7$ and the final
estimate,
\begin{equation}
\delta v = 3.2\times 10^{-5}~{\rm cm~s}^{-1}\left(F\over
2.7\right)\left({\langle \vert v_a^2\vert \rangle^{1/2}\over 20~{\rm
    km~s^{-1}}}\right)~.
\label{newt}
\end{equation}

The PTA observatories are located on Earth but the differential effect
between the Earth and the Sun is suppressed by more than an order of
magnitude relative to the jitter of the Sun as a result of a spatial
tidal factor $\sim (r_{\rm E}/r_a)^3$ as well as an orbital
time-averaging factor, because the Earth-Sun separation, $r_{\rm E}$,
is a few times smaller than orbital radii of the perturbing asteroids,
$r_a$. Therefore, it is sufficient to consider the Brownian motion of
the Sun which is tracked to zeroth order by the Earth, as done here.

For semi-major axes in the range of 2-4~au, the characteristic
frequency of the perturbations would be of order the typical orbital
frequency of the perturbing asteroids, $f\sim (3~{\rm
  yr})^{-1}=10^{-8}~{\rm Hz}$.

The unmodeled random walk of the Sun's velocity vector in 3D is
characterized by a correlation time of $f^{-1}\sim 3~{\rm yr}$. It
leads to Doppler-shift fluctuations in pulsar timing that is
correlated among different pulsars. The resulting Doppler vector
fluctuations with a stochastically varying orientation, could
mascarade as a stochastic gravitational wave background of a
characteristic strain,
\begin{equation}
h_c\sim \left({\delta v\over c}\right)\sim 1.1\times 10^{-15} \left(F\over
2.7\right)\left({\langle \vert v_a^2\vert \rangle^{1/2}\over 20~{\rm
    km~s^{-1}}}\right)~,
\label{hc}
\end{equation}
which makes up to a $\sim 20\%$ contribution to the amplitude to the
gravitational wave signal at $\sim 10^{-8}~{\rm Hz}$ reported by
PTAs~\citep{Ant2022,Agazie2023,Reardon2023}.

Each individual asteroid induces by itself a much smaller Solar velocity
fluctuation of order~\citep{BT},
\begin{equation}
{\Delta v_\odot \over c} \sim \left({2Gm_a\over
  b_a v_a c}\right) \sim 5\times 10^{-17}\left({m_a\over 10^{21}~{\rm g}}\right)
\left({b_a\over 3~{\rm au}}\right)^{-1}~,
\label{imp}
\end{equation}
where $b_a$ is its distance of closest approach to the Sun (perihelion). 

At any given time, the distribution of unmodeled or improperly
mass-calibrated asteroids would be randomly lopsided towards one
hemisphere around the Sun, with a Poisson excess of $\sim
(N_a/2)^{1/2}$ asteroids. Indeed, this enhancement yields the larger
cumulative signal expressed in equation~(\ref{hc}). The direction and
amplitude of the Poisson fluctuations accelerating the Sun would
change randomly over asteroid orbital times in the range, $\sim
10$--$3~{\rm yr}$, corresponding to the frequency range of $f\sim
3$--$10~{\rm nHz}$.

\section{Discussion}

I have shown that unmodeled or improperly mass-calibrated
asteroids result in a Brownian motion of the Sun relative to the Solar
system barycenter with a velocity dispersion given by
equation~(\ref{newt}). This leads to a non-negligible noise for
correlated timing residual of pulsars at an amplitude (\ref{hc}) and
frequencies relevant to PTAs~\citep{Ant2022,Agazie2023,Reardon2023}.

When squaring the right-hand-side of equation (\ref{mom}), we ignored
correlations between the velocities of different unmodeled
asteroids. Any such correlations as a result of the phase-space
clustering of asteroids would enhance the expected noise. For example,
there are more than 100 Jupiter's Trojans with diameters in the range
40-80km exceeding $10^{22}$g in total mass, and there are many known
families in the Main asteroid belt (such as Flora, Eunomia, Eos,
Hungaria, Karin, Koronis, Phocaea, Vesta and Themis; see~\citet{Lema})
in addition to the Trojans of Earth and Mars. Planetary Trojans might
already be accounted for in the mass budget uncertainty of the
corresponding planets. But for families with other orbital
periods~\citep{Nes1}, the collective mass of a perturber will be that
of a physical cluster of asteroids rather than a single asteroid. The
additional effect of these clusters is an enhancement of the noise
amplitude in equation~(\ref{nomom}), which can be accounted for by
using the cluster mass and number in place of $m_a$ and $N_a$. This
enhancement cannot exceed an extra factor of $\sqrt{N_a}$ relative to
the independent asteroids account of equation (\ref{nomom}), as this
extra factor saturates the upper limit associated with placing all
asteroids in a single cluster.

An increase in the number of Main-belt asteroids included in the Solar
system ephemeris~\citep{Eph} and more accurate mass and orbit
determinations of asteroids by the Legacy Survey of Space and Time
(LSST) of the Vera C. Rubin Observatory in Chile~\citep{LS}, would
allow better modeling of the noise they introduce to PTAs.

\bigskip
\bigskip
\bigskip
\bigskip
\section*{Acknowledgements}

I thank two anonymous referees for excellent suggestions, Michael
Fedderke for correcting a typo and sharing insights, and Matt Holman
and Peter Veres for helpful comments. This work was supported in part
by Harvard's {\it Black Hole Initiative}, which is funded by grants
from JFT and GBMF.

\bigskip
\bigskip
\bigskip

\bibliographystyle{aasjournal}
\bibliography{t}
\label{lastpage}
\end{document}